\PassOptionsToPackage{unicode}{hyperref}
\PassOptionsToPackage{hyphens}{url}
\PassOptionsToPackage{dvipsnames,svgnames,x11names}{xcolor}
\documentclass[12pt]{article}

\usepackage{graphicx}%
\usepackage{multirow}%
\usepackage{amsmath,amssymb,amsfonts}%
\usepackage{amsthm}%
\usepackage{mathrsfs}%
\usepackage[title]{appendix}%
\usepackage{xcolor}%
\usepackage{textcomp}%
\usepackage{manyfoot}%
\usepackage{booktabs}%
\usepackage{algorithm}%
\usepackage{algorithmicx}%
\usepackage{algpseudocode}%
\usepackage{listings}%

\usepackage{amsmath,amssymb}
\usepackage{iftex}
\ifPDFTeX
  \usepackage[T1]{fontenc}
  \usepackage[utf8]{inputenc}
  \usepackage{textcomp} 
\else 
  \usepackage{unicode-math}
  \defaultfontfeatures{Scale=MatchLowercase}
  \defaultfontfeatures[\rmfamily]{Ligatures=TeX,Scale=1}
\fi
\usepackage{lmodern}
\ifPDFTeX\else  
\fi
\IfFileExists{upquote.sty}{\usepackage{upquote}}{}
\IfFileExists{microtype.sty}{
  \usepackage[]{microtype}
  \UseMicrotypeSet[protrusion]{basicmath} 
}{}
\makeatletter
\@ifundefined{KOMAClassName}{
  \IfFileExists{parskip.sty}{%
    \usepackage{parskip}
  }{
    \setlength{\parindent}{0pt}
    \setlength{\parskip}{6pt plus 2pt minus 1pt}}
}{
  \KOMAoptions{parskip=half}}
\makeatother
\usepackage{xcolor}
\makeatletter
\ifx\paragraph\undefined\else
  \let\oldparagraph\paragraph
  \renewcommand{\paragraph}{
    \@ifstar
      \xxxParagraphStar
      \xxxParagraphNoStar
  }
  \newcommand{\xxxParagraphStar}[1]{\oldparagraph*{#1}\mbox{}}
  \newcommand{\xxxParagraphNoStar}[1]{\oldparagraph{#1}\mbox{}}
\fi
\ifx\subparagraph\undefined\else
  \let\oldsubparagraph\subparagraph
  \renewcommand{\subparagraph}{
    \@ifstar
      \xxxSubParagraphStar
      \xxxSubParagraphNoStar
  }
  \newcommand{\xxxSubParagraphStar}[1]{\oldsubparagraph*{#1}\mbox{}}
  \newcommand{\xxxSubParagraphNoStar}[1]{\oldsubparagraph{#1}\mbox{}}
\fi
\makeatother

\usepackage{longtable,booktabs,array}
\usepackage{calc} 
\usepackage{etoolbox}
\makeatletter
\patchcmd\longtable{\par}{\if@noskipsec\mbox{}\fi\par}{}{}
\makeatother
\IfFileExists{footnotehyper.sty}{\usepackage{footnotehyper}}{\usepackage{footnote}}
\makesavenoteenv{longtable}
\usepackage{graphicx}
\makeatletter
\def\maxwidth{\ifdim\Gin@nat@width>\linewidth\linewidth\else\Gin@nat@width\fi}
\def\maxheight{\ifdim\Gin@nat@height>\textheight\textheight\else\Gin@nat@height\fi}
\makeatother
\setkeys{Gin}{width=\maxwidth,height=\maxheight,keepaspectratio}
\makeatletter
\def\fps@figure{htbp}
\makeatother

\makeatletter
\@ifpackageloaded{caption}{}{\usepackage{caption}}
\AtBeginDocument{%
\ifdefined\contentsname
  \renewcommand*\contentsname{Table of contents}
\else
  \newcommand\contentsname{Table of contents}
\fi
\ifdefined\listfigurename
  \renewcommand*\listfigurename{List of Figures}
\else
  \newcommand\listfigurename{List of Figures}
\fi
\ifdefined\listtablename
  \renewcommand*\listtablename{List of Tables}
\else
  \newcommand\listtablename{List of Tables}
\fi
\ifdefined\figurename
  \renewcommand*\figurename{Figure}
\else
  \newcommand\figurename{Figure}
\fi
\ifdefined\tablename
  \renewcommand*\tablename{Table}
\else
  \newcommand\tablename{Table}
\fi
}
\@ifpackageloaded{float}{}{\usepackage{float}}
\floatstyle{ruled}
\@ifundefined{c@chapter}{\newfloat{codelisting}{h}{lop}}{\newfloat{codelisting}{h}{lop}[chapter]}
\floatname{codelisting}{Listing}

\makeatother
\makeatletter
\@ifpackageloaded{caption}{}{\usepackage{caption}}
\@ifpackageloaded{subcaption}{}{\usepackage{subcaption}}
\makeatother

\ifLuaTeX
  \usepackage{selnolig}  
\fi
\usepackage[]{natbib}
\usepackage{bookmark}

\IfFileExists{xurl.sty}{\usepackage{xurl}}{} 
\hypersetup{
  pdftitle={Title},
  pdfauthor={Author 1; Author 2},
  pdfkeywords={3 to 6 keywords, that do not appear in the title},
  colorlinks=true,
  linkcolor={blue},
  filecolor={Maroon},
  citecolor={Blue},
  urlcolor={Blue},
  pdfcreator={LaTeX via pandoc}}

\newcommand{\anon}{1}

\begin{document}

\def\spacingset#1{\renewcommand{\baselinestretch}%
{#1}\small\normalsize} \spacingset{1}


\if1\anon
{
  \title{\bf Efficient model exploration with the integrated nested Laplace approximation}
  \author{H\'ector L\'opez-G\'omez\thanks{
    This work has been supported by grant PID2022-136455NB-I00, funded by MCIN/AEI/10.13039/501100011033/FEDER and the European Regional Development Fund, grant CIAICO/2022/165, funded by Direcci\'on General de Ciencia e Investigaci\'on (Generalitat Valenciana, Spain), and grant SBPLY/21/180501/000241, funded by Consejer\'ia de Educaci\'on, Cultura y Deportes (Junta de Comunidades de Castilla-La Mancha, Spain) and FEDER.}\hspace{.2cm}\\
    Department of Mathematics, School of Industrial Engineering-Albacete,\\
    Universidad de Castilla-La Mancha\\
    and \\
    Virgilio G\'omez-Rubio \\
    Department of Mathematics, School of Industrial Engineering-Albacete,\\
    Universidad de Castilla-La Mancha\\
    and\\
    Gonzalo Garc\'ia-Donato\\
    Department of Economy and Finance, Faculty of Economics,\\
    Universidad de Castilla-La Mancha
    }
  \maketitle
} \fi

\if0\anon
{
  \bigskip
  \bigskip
  \bigskip
  \begin{center}
    {\LARGE\bf Efficient model exploration with the integrated nested Laplace approximation}
\end{center}
  \medskip
} \fi

\bigskip
\begin{abstract}
Model and variable selection are important topics in Bayesian inference. In particular, the selection of different fixed, random effects and hyperparameters in hierarchical models can be difficult because of their complex structure. In this paper, we introduce the use of approximate Bayesian inference for model and variable selection for hierarchical Bayesian models.

The approach is based on the application of Markov chain Monte Carlo methods on the model space. In particular, the Metropolis-Hastings algorithm is used to sample from the set of model indices. In this way, the marginal likelihood is used to compute the acceptance probability so that it is not required to estimate the parameter models directly. For this, the integrated nested Laplace approximation (INLA) is used because it provides accurate estimates of the marginal likelihood and it can also provide estimates of the posterior marginal of the model parameters.

This method can be applied not only to variable selection but to a wide range of problems subject to model uncertainty. To illustrate the potential of this approach, examples on variable selection, changepoint models and log-Gaussian Cox processes are developed. 

\end{abstract}

\noindent%
{\it Keywords:} INLA, Markov chain Monte Carlo, model selection, variable selection
\vfill

\newpage
\spacingset{1.8} 

\section{Introduction}
Model and variable selection are challenging problems in Bayesian inference. There is extensive literature on variable selection \citep[for example,][]{TadesseVannucci:2021}.
In the context of Bayesian hierarchical models, it is also important to address the selection of models with mixed effects. This is particularly important for highly-structured hierarchical models such as spatio-temporal or joint models.

An early attempt to address these topics was provided by \cite{CarlinChib:1995}, that relied on Markov chain Monte Carlo methods to explore the model space, compute the posterior probability of each model and estimate their respective parameters. They rely on a Gibbs sampler that iterates across the space of models as well as the respective parametric spaces. However, this approach requires the use of pseudo-priors for the case in which samples from parameters for non-selected models during Gibbs sampling.  

Similarly,  \cite{Green:1995} proposed a reversible jump Markov chain Monte Carlo (RJMCMC) algorithm, so that both model selection and parameter estimation can be done at the same time. RJMCMC introduces a new type of movement in Markov chain Monte Carlo that allows jumping from one model to another in a consistent way with the change of dimension in the parametric space. When jumps between one model and another occur, the parametric space will most likely change \citep[see, for example,][]{RichardsonGreen:1997}. This means that the dimension of the parametric space will also change and this situation needs to be taken care of accordingly. This makes RJMCMC more complex than traditional MCMC approaches that focus on the estimation of the posterior distribution of the model parameters but provides a comprehensive framework for model estimation and selection.

In this paper, we propose an approach based on MCMC to conduct variable and model selection.  Once the full model space is defined, the proposed algorithm uses Metropolis-Hastings (M-H) to explore the model space and compute the posterior probability of each model. This approach produces a discrete Markov chain on the set of indices of the models. During the M-H algorithm, a proposal distribution will propose new models to explore by taking the current one and modifying its structure slightly. For example, in the case of variable selection, new proposals can be obtained by adding or removing a covariate in the linear predictor. More complex models may require different strategies to explore the model space. 



At each step of the Metropolis-Hastings procedure it is necessary to fit the proposed model and estimate its marginal likelihood. The integrated nested Laplace approximation  \citep[INLA,][]{Rueetal:2009} is used in this case as it provides a faster way to fit models than traditional MCMC methods and also calculates the marginal likelihood of the model, which is required when computing the acceptance probability of the proposed model.

This is similar in nature to the INLA with MCMC approach \cite{GomezRubioRue:2018}. Note that this requires computing the marginal likelihood of the proposed model at each step of the Metropolis-Hastings procedure. However, in this particular case we can take advantage of the fact that the model space is discrete so that models are only fitted the first time they are proposed. Furthermore, as a by-product of using INLA to estimate the marginal likelihood, the posterior marginal of the model parameters will also be obtained.


This paper is structured as follows. First, Bayesian inference for model choice is discussed in Section~\ref{sec:Bayesian}. Next, the integrated nested Laplace approximation is described briefly. In Section~\ref{sec:INLA}, the methodology used for model selection with Markov chain Monte Carlo is described. In Section~\ref{sec:examples}, we include a number of examples to illustrate the proposed methodology in linear regression, change-point models and spatial models. Finally, in Section~\ref{sec:discussion} we provide final remarks and a discussion.

\section{Bayesian inference for model choice}
\label{sec:Bayesian}

Let us consider a set of $M$ (parametric) models $\Omega_M = \{\mathcal{M}_i\}_{i=1}^M$.
These models are often used to represent how data have been generated in a particular problem. A typical example is linear regression in which a response variable is modeled using a set of covariates. In this particular case, the set of models represents the different ways in which covariates can be included in the model.

Bayesian inference about the different models involves computing the posterior
probability of each model $\pi(\mathcal{M}_i \mid \mathcal{D})$, where $\mathcal{D}$ represents the observed data. The posterior probability can be expressed as follows using Bayes's rule:

$$
\pi(\mathcal{M}_i \mid \mathcal{D}) = \frac{\pi(\mathcal{D} \mid \mathcal{M}_i) \pi(\mathcal{M}_i)}{\pi(\mathcal{D})}
$$
\noindent
Here, $\pi(\mathcal{D} \mid \mathcal{M}_i)$ is the probability of observing the data given model $\mathcal{M}_i$ (also known as marginal likelihood) and $\pi(\mathcal{M}_i)$ the prior probability of model $i$. Term $\pi(\mathcal{D})$ is the probability of the data (regardless of the model), which is often difficult to compute. This term can also be regarded as a normalizing constant so that the sum of all posterior probabilities, $\sum_{i=1}^M \pi(\mathcal{M}_i \mid \mathcal{D})$, is equal to one.

Regarding the prior distribution of the models, a simple approach is to take
 $\pi(\mathcal{M}_i) = 1 / M,\ i=1,\ldots, M$, so that all models have the same probability a priori. Other approaches may involve penalizing by the complexity of the model (e.g., the number of parameters) so that more complex models tend to have a smaller probability a priori \citep[see, for example,][]{GomezRubioetal:2021}.

In practice, computing the marginal likelihood $\pi(\mathcal{D} \mid \mathcal{M}_i)$ involves estimating the parameters $\theta^{(i)} = \left(\theta^{(i)}_1,\ldots,\theta^{(i)}_{P_i}\right)$ that define model $\mathcal{M}_i$, i.e., computing the posterior distribution $\pi(\theta^{(i)} \mid \mathcal{D})$. Note that the previous distribution involves conditioning on model $\mathcal{M}_i$, i.e., $\pi(\theta^{(i)} \mid \mathcal{D}) \equiv \pi(\theta^{(i)} \mid \mathcal{D}, \mathcal{M}_i)$. The corresponding posterior marginal distributions will be represented by 
$ \pi(\theta^{(i)}_j \mid \mathcal{D}),\ j=1,\ldots,P_i$. For simplicity, when the model is clearly stated by the context, we will simply write $\pi(\theta \mid \mathcal{D})$ and $\pi(\theta_j \mid \mathcal{D})$, so that the upper-index as well as the conditioning model are omitted (i.e., model $\mathcal{M}_i$ is omitted).

The marginal likelihood can then be computed as

$$
\pi(\mathcal{D} \mid \mathcal{M}_i) = \int_{\theta\in\Theta_i} \pi(\mathcal{D} \mid \theta) \pi(\theta) \mathrm{d} \theta\ \ i=1,\ldots,M .
$$ 
\noindent
Here, $\Theta_i$ represents the parametric space of the parameters of model $\mathcal{M}_i$.

In practice, computing the marginal likelihood is a difficult problem. However, when Markov chain Monte Carlo are used for inference, the marginal likelihood can be computed from the sample obtained \citep[see, for example,][]{Chib:1995,ChibJeliazkov:2001,Vitoratouetal:2014}.

Note also that computing the marginal likelihood is key for comparing models using the Bayes factor \citep{KassRaftery:1995}. When comparing model $\mathcal{M}_k$ and $\mathcal{M}_l$ (with $k,l\in\{1,\ldots,M\})$, the Bayes factor is computed as

$$
\mathrm{BF}(\mathcal{M}_k, \mathcal{M}_l)  = 
\frac{\pi(\mathcal{D} \mid \mathcal{M}_k)}{\pi(\mathcal{D} \mid \mathcal{M}_l)} =
\frac{\pi(\mathcal{M}_k \mid \mathcal{D})}{\pi(\mathcal{M}_l \mid \mathcal{D})}  \frac{\pi(\mathcal{M}_l)}{\pi(\mathcal{M}_k)},\ k,l\in\{1,\ldots,M\}.
$$
\noindent
In a nutshell, the Bayes factor is a measure of how the data supports model  $k$ compared to model $l$ penalized by the complexity. Note that when all models have the same prior probability the Bayes factor reduces to the ratio of the posterior probabilities.

As discussed above, the posterior probabilities of the models can be used for model selection. In this context, the preferred model should be the one with the highest posterior probability and inference on the model parameters will rely upon the posterior distribution of this particular model.

Alternatively, instead of selecting a single model, inference could be based on the average of the marginals estimated for each of models so that the posterior distribution of the models act as weights. This is known as Bayesian model averaging \citep{Rafteryetal:1997}. Note that it is likely that not all parameters considered across models are estimated with all the models. This means that only those models in which a parameter appears will be used for model averaging (and their weights re-scaled accordingly).

Hence, conducting inference about model adequacy requires estimating the posterior distribution of the model parameters as well as computing its posterior probability. When the number of models is large, estimating all possible models is unfeasible or simply impossible. For example, when considering variable selection, the number of possible predictors may be very large and computationally efficient approaches are required \citep{FanLv:2009}.

\section{Makov chain Monte Carlo for Model Choice}
\label{sec:MCMC}

Exploring the model space $\Omega_{\mathcal{M}}$ can be challenging. In particular, listing all possible models could be impossible when the number of models is large. In this regard, a Monte Carlo approach could be followed in order to propose the models to explore. Notice that estimating the posterior distributions of the model parameters for a given model is of secondary interest. This is so because it is likely that only a handful of models will actually have a high posterior probability and, hence, be of interest. 

As stated by \cite{CarlinChib:1995}, implementations of MCMC for models with varying structure can be difficult as it violates some of the sufficient requirements for convergence. Following their work, inference will be based on the index that represents the models, so that samples from this index will be obtained. They propose Gibbs sampling to estimate both the posterior model probabilities and posterior distribution of the model parameters. As summarized above, \cite{Green:1995} proposes a more general framework for MCMC to estimate the posterior model probabilities and the posterior distribution of the  model parameters. In both cases, the estimation algorithm require sampling from the parametric space.

%
%
%

The new approach proposed in the following only requires estimating the marginal likelihood and does not require
sampling from the space of parameters. In particular, the Metropolis-Hastings algorithm is used in order to explore the model space $\Omega_{\mathcal{M}}$. Hence, it is required to define a proposal distribution $q(\cdot \mid \cdot)$ that proposes new models given the current one. The required models will be fit with the integrated nested Laplace approximation, that provides a fast way of fitting mixed-effects models with different structures \cite[see, for example,][]{GomezRubio:2020}.

The Metropolis-Hastings for exploring the model space is as follows:

\begin{enumerate}

\item Set $i_{0}$ to be the index of the initial model $\mathcal{M}_{i_0}$.

\item For $j=1$ to $n_{iter}$ do
\begin{enumerate}
\item Sample the index $i^*$ of a proposed model from $q(\cdot\mid i_{j-1})$.

\item Compute acceptance probability $\alpha$:

$$
\alpha = \min\left\{1, \frac{q(i_{j-1} \mid i^*)\pi(\mathcal{D} \mid \mathcal{M}_{i^*})\pi(\mathcal{M}_{i^*})}{q(i^* \mid i_{j-1})\pi(\mathcal{D} \mid \mathcal{M}_{i_{j-1}})\pi(\mathcal{M}_{i_{j-1})}}\right\}
$$

\item Sample value $u_j$ from uniform distribution in $(0,1)$.

\item If $u_j < \alpha$, then $i_j=i^*$. Otherwise, $i_j=i_{j-1}$.
\end{enumerate}

\item Compute posterior probabilities from samples. 

\end{enumerate}

A number of remarks about the former algorithm are needed. First of all, the Metropolis-Hastings algorithm requires the prior probability, the marginal likelihod of each proposed model, as well as the probabilities associated to the proposal distribution. The prior probabilities and the proposal probabilities are known as they are defined by the user. The marginal likelihood can be challenging to obtain but it can be computed for a few of models. Alternatively, it can be approximated using a number of techniques and we propose the use of the integrated nested Laplace approximation.

The proposal distribution will need to be tailored to each specific case. A naïve approach could be to simply propose a model at random from the model space. However, this will most likely perform poorly. Next, we provide some guidelines for model selection and propose some general guidelines for the more general case. Furthermore, in the examples in Section~\ref{sec:examples} we have developed a number of case studies in which different proposal distributions are used depending on the particular framework.

\subsection{Estimation of the marginal likelihood}
\label{sec:INLA}

Bayesian inference will be conducted using approximate inference with the integrated nested Laplace approximation \citep[INLA,][]{Rueetal:2009}. The INLA method provides a fast way to approximate the posterior marginals of the model parameters and latent effects for a large class of mixed-effects models. Furthermore, INLA can produce reliable approximations of other quantities of interest such as the marginal likelihood of a model.

Specifically, INLA will be used to obtain accurate estimates of the marginal likelihood. \cite{HubinStorvik:2016b} have conducted a thorugh study and have concluded that the estimates provided by INLA are acccurate enough. In addition, \cite{GomezRubioRue:2018} have used these approximations to implement Markov chain Monte Carlo methods using the estimates of the marginal likelihood provided by INLA sucessfully.

This approximation of the marginal likelihood for a given model $\mathcal{M}_i$ is obtained by computing

$$
\tilde\pi(\mathcal{D} \mid \mathcal{M}_i) =
\int \frac{\pi(\theta,\mathbf{x}, \mathcal{D})}{\pi(\mathbf{x} \mid \mathbf{\theta}, \mathcal{D})}\Bigg|_{\mathbf{x}=\mathbf{x}^*(\mathbf{\theta})} d\mathbf{\theta}
$$
\noindent
Here, $\mathbf{x}$ represents the vector of latent effects and $\mathbf{x}^*(\mathbf{\theta})$ refers to the posterior mode of the vector of latent effects $\mathbf{x}$ for a given value of the vector of hyperparameters $\mathbf{\theta}$. Note that $\mathbf{x}$ and $\mathbf{\theta}$ will have different dimensions depending on $\mathcal{M}_i)$ but that the $i$ index has been ommited for simplicity.

Finally, the other posterior marginals provided by INLA are those for the hyperparameters, $\pi(\theta_{\bullet} \mid \mathcal{D})$, and those of the random effects,  $\pi(\mathbf{x}_{\bullet} \mid \mathcal{D})$. Note that these are a by-product of model fitting with INLA as our primary interest is in computing the marginal likelihoods. However, these marginals will be of interest for the models with the largest posterior probabilities.

\section{Strategies to explore the model space}

The proposal distribution that appears in the Metropolis-Hastings algorithm can be regarded as how the model space is explored. Hence, given a model, a neighbourhood of models could be defined around it and one of them picked up at random. In a general framework, it is necessary to think about how proposals could be made. The proposal distribution should propose new models that are "close" to the current one. Proposing similar models is important to have some sense of direction when exploring the model space and to avoid a situation in which models are proposed almost completely at random. 

For this reason, when defining the proposal distribution, it is important to do it in a way that proposes smooth changes in the current model to avoid models that are too different from each other and explore the model space parsimoniously. In this regard, it is important to determine some structural characteristics in the model to be able to define when two models are close or similar. 

Next, we consider the particular case of variable selection. A simple approach is to consider that two models are neighbours when they differ in a single variable in the linear predictor. However, models differing in two or more variables could be proposed to produce longer jumps. In the examples in Section~\ref{sec:examples}, we propose different types of models in which the definition of the proposal distribution is discussed.

\subsection{Variable selection}
\label{sec:varsel}

For the particular case of variable selection, the space of models is defined by all possible models that appear depending on whether a particular variable is included or not. When $p$ different predictors are available, the total number of models is $2^p$. Hence, even for a moderately small number of predictors, the model space can be quite large and difficult to explore.

To illustrate the use of our algorithm for variable selection, we propose a simple approach to propose a new model from the current one. First of all, it is decided whether a variable model is removed or, alternatively, a variable not in the model is included. This is decided at random, with 0.5 probability for each action. If a variable in the model is removed, then one is selected at random and the model proposed is one with the same variables in the current model plus the selected predictor. Similarly, if a variable is removed from the current model, then one of them is selected at random. The proposed model is the one with all the variables in the current one but the selected predictor.

Note that there are two edge-cases in this approach. First of all, if the model only contains the intercept, the only action is to propose a new model with a new variable added at random. Secondly, for the full model, the only possible action is to propose a model in which one of the variables is removed at random.

Given a model with $k$ predictors, with $k\in\{1, p-1\}$, the probabilities associated to the proposal
distribution are the following:

\begin{itemize}
    \item Removing a variable at random: $q(\cdot \mid \cdot) = \frac{1}{2}\frac{1}{k}$.
    \item Adding a variable at random: $q(\cdot \mid \cdot) = \frac{1}{2}\frac{1}{p-k}$.
\end{itemize}

Acceptance probabilities can be obtained in closed form in this setting. Assuming that all models have the same prior probability, acceptance probabilities take the following forms:

\begin{itemize}
    \item Removing a variable at random: $\alpha=\min\left\{1, \frac{p-k+1}{k}\frac{\pi(\mathcal{D} \mid \mathcal{M}_{i^*})}{ \pi(\mathcal{D} \mid \mathcal{M}_{i_{j-1}})}\right\}$.
    
    \item Adding a variable at random: $\alpha=\min\left\{1, \frac{k+1}{p-k}\frac{\pi(\mathcal{D} \mid \mathcal{M}_{i^*})}{\pi(\mathcal{D} \mid \mathcal{M}_{i_{j-1}})}\right\}$.

\end{itemize}

Although this approach to propose new models is simple, it is also very fast to compute. In particular, once a model has been fit, it is not necessary to fit it again in order to compute its marginal likelihood.

The way in which new models are proposed could be modified in different ways to gain efficiency. For example, instead of adding or removing a variable from the model completely at random, variables could be proposed for selection with different probabilities depending on how they are related with the outcome variable. This can be done by conducting a pre-screening to fit univariate models and determine the variables which are more likely to explain the outcome best. For example, the marginal likelihoods can be computed and res-scaled to sum up to one and then use these probabilities to introduce a variable in the model instead of assuming equal probabilities.

\subsection{Mixed-effects and hierarchical models}
\label{subsec:GS}

\cite{Xuetal:2023} discuss the Bayesian model selection for generalized mixed-effects models. For the case of mixed-effects models, we will focus on the selection of random effects and keep the same fixed effects. This is so because confounding effects may appear between fixed and random effects. For example, this is not uncommon in models with spatially correlated random effects \cite{Urdangarinetal:2023}.

Model selection for hierarchical models may involve choosing among a number of components in the model such as, for example, different structures for the random effects. In this regard, we follow the Gibbs sampling algorithm for Bayesian variable selection proposed by several authors \citep{GeorgeMcCulloch:1993,GDMB:2013,GarciaDonatoSteel:2021}. First of all, each model is represented by a binary vector $\gamma$ that identifies the structures present in the model. The algorithm can start at, say, $\gamma_0$ with all its elements equal to zero, which would represent a baseline model as none of the structures is in the model.

At iteration $j$ of the Gibbs sampling algorithm, given a binary vector $\gamma_{(j-1)}=(\gamma_{1(j-1)}, \gamma_{2(j-1)}, \ldots, \gamma_{p(j-1)})$ (from the previous step), each element $\gamma_{k(j-1)}$ is replaced in turn by $1-\gamma_{k(j-1)}$ and this change accepted with probability 

$$
\frac{\pi(\mathcal{D} \mid \gamma_*)\pi(\gamma_*)}{\pi(\mathcal{D} \mid \gamma_*)\pi(\gamma_*)+\pi(\mathcal{D} \mid \gamma_{j-1})\pi(\gamma_{j-1})}
$$
Here, $\gamma_*$ represents the resulting proposed model after shifting some of the components in $\gamma_{(j-1)}$.

Once all elements have been processed, some of them may have been shifted. Then, $\gamma_{j}$ is set to the resulting binary vector after processing all its elements and another iteration is conducted.

In this case, the priors should penalize more complex models, i.e., models with a larger number of structures. In particular, for a model with $k$ components among a total of $n_C$ components, we follow \cite{ScottBerger:2010} and use of the following prior:

$$
\pi_1(\gamma) = \frac{1}{(1 + n_C){n_C \choose k}}
$$

Note that this will assign the same prior probability to the model with no covariates and the full model, that is, the prior will not penalize the increasing model complexity appropriately. As an alternative prior, a truncated prior can be set by pretending that there are $2n_C$ covariates and truncating it at $n_C$ variables, i.e.,

$$
\pi_2(\gamma) \propto \frac{1}{(1 + 2n_C){2n_C \choose k}},\ k=0,\ldots, n_C
$$
\noindent
This will ensure a decaying prior with the number of components in the model from 0 to $n_C$ components.

\section{Examples}
\label{sec:examples}

In this section a number of examples are developed to illustrate the previous methodology. First of all, an example is developed in Section~\ref{subsec:example1} with the aim of describing the proposed methodology for variable selection. Section~\ref{subsec:changepoint} provides an example of fitting a change-point model. Finally, Section~\ref{subsec:example3} develops a real application of the methodology to model selection of log-Gaussian Cox processes.

\subsection{Variable selection}
\label{subsec:example1}

In order to provide an example of variable selection that is more challenging and still manageable for comparison purposes, we will analyze the \texttt{mtcars} dataset available in the R software for statistical computing. This dataset includes data about fuel consumption and 10 variables with aspects of automobile design. All variables have been standardized. Hence, the aim is to fit a model to explain fuel consumption on these 10 variables. Together, the total number of possible models is $2^{10}=1024$, which poses a challenge. 

Models will be indexed by their binary representation of the included covariates, e.g., if covariates $x_2$, $x_5$ and $x_{10}$ are included in the model, it index is 0100100001 (which is model number 289). Given a model $\mathcal{M}_m$ (with $m=0,\ldots,1023$), we can define an index set $\mathcal{I}_m \subseteq \{1,\ldots, 10\}$ with the indices of the variables in the model. Hence, models can be defined in the following way:

\begin{eqnarray}
y_i &\sim &N(\mu_i,\tau) \nonumber\\
\mu_i & = & \alpha+\sum_{j \in \mathcal{I}_m} \beta_j x_{ij} \nonumber
\label{eq:mtcars}
\end{eqnarray}
 
Priors for all the fixed effects (i.e., intercept and covariates) are Gaussian with zero mean and a precision of 1 for the intercept and the coefficients of the covariates, and a Gamma with parameters 0.00001 and 0.00001 for the precision $\tau$. Note that all the variables have been standardized and that these priors make sense. When both the dependent variable and the regressors are scaled to have unitary variance, using independent $\beta_j\sim N(0, 1)$ is a natural independent approximation to the $g$--prior \citep{Zellner:1986}. This specification preserves the scale--invariance properties of the original $g$--prior while yielding independent coefficients. Independency among the regressors is a convenient assumption that is frequently assumed in the literature, as in \cite{GeorgeMcCullogh:1997}. Finally, a uniform prior is considered for the models themselves. 

Variable selection was conducted using the Metropolis-Hastings algorithm described in Section~\ref{sec:MCMC}. In addition, the 1024 different models were fit and their posterior probabilities computed. In all cases, all models have the same prior probabilities. i.e., $\pi(\mathcal{M}_m)=1 / 1024,\ m=1,\ldots,1024$.

\begin{figure}[h!]
\caption{Models fit to the \texttt{mtcars} dataset with a posterior probability greater than 0.005.}
\label{fig:res-mtcars}
\includegraphics{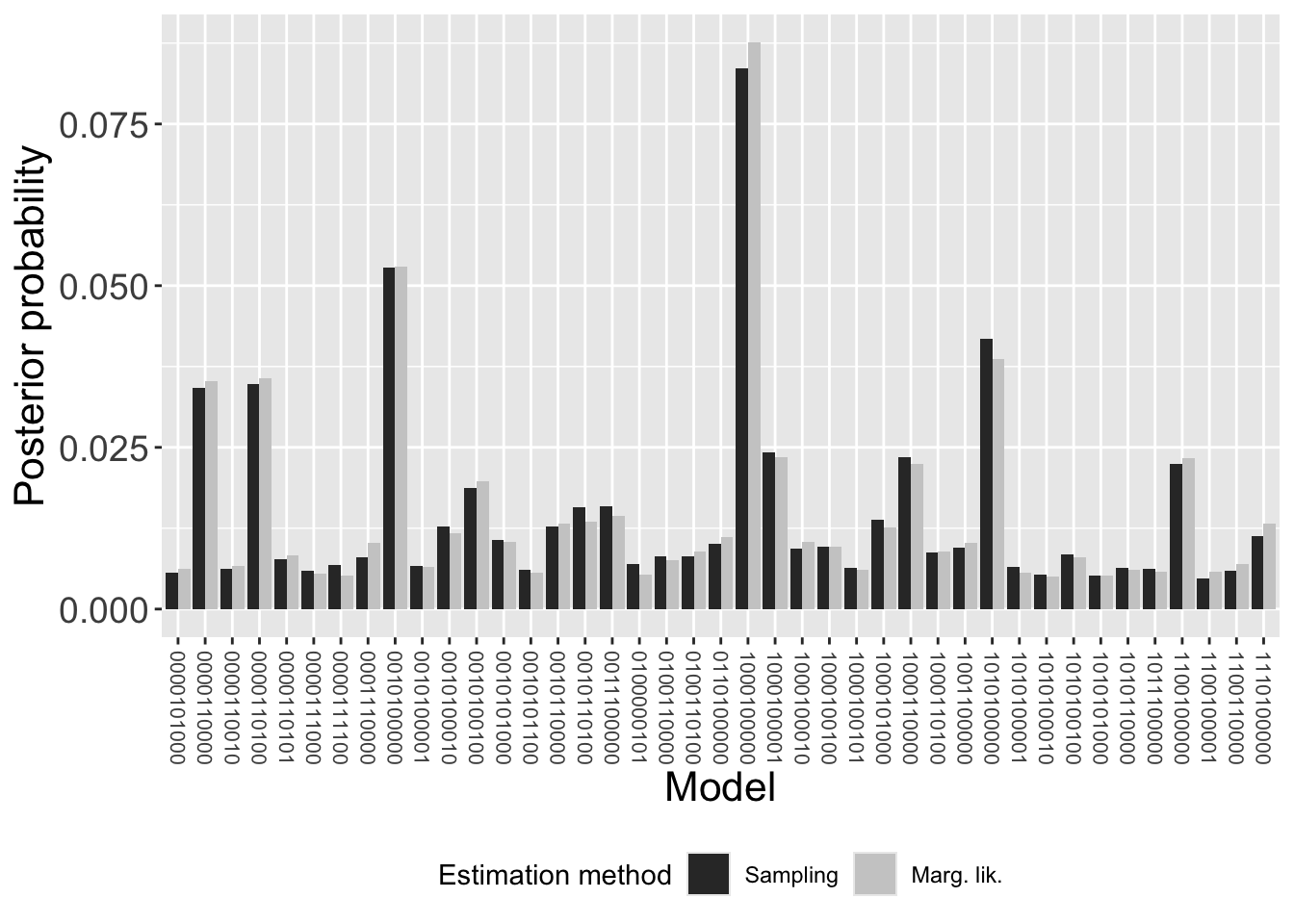}
\end{figure}

Most models show a posterior probability very close to 0. The 41 models that have a posterior probability larger than 0.005 are shown in Figure~\ref{fig:res-mtcars}. These posterior probabilities have been computed by using the MCMC output as well as the marginal likelihood after fitting all 1024 models. It is worth noting that the model space has been extensively explored and that only 36 (out of 1024) have been left unexplored.

Similarly, Table~\ref{tab:res-mtcars} shows the estimates of the model parameters, which have been computed by conducting a Bayesian model averaging with all models. In the case of the coefficients of the covariates, only the models in which they are included have been considered when doing the BMA and the posterior probabilities of the models have been re-scaled to sum up to one. Furthermore, the posterior probability of inclusion for each covariate is available. These can be computed by simply adding the posterior probabilities of the models in which each variable is included.

\addtolength{\tabcolsep}{-0.3em}
\begin{table}[h!]
\caption{Summary statistics of the model parameters for the \texttt{mtcars} dataset.}
\label{tab:res-mtcars}
\centering
\begin{scriptsize}
\begin{tabular}{ll|c|c|cc|c|c|c|cc|c|} 
& & \multicolumn{5}{c|}{MCMC} & \multicolumn{5}{c|}{Marg. lik.} \\
Variable & Acronym & Mean & St. dev. & \multicolumn{2}{c|}{95\% C.I.} & Prob. inclusion &
Mean & St. dev. & \multicolumn{2}{c|}{95\% C.I.} & Prob. inclusion\\ 
\hline
Number of cylinders & cyl & -0.347 & 0.209 & -0.730 & 0.103 & 0.469 & -0.348 & 0.210 & -0.732 & 0.104 & 0.474\\
Displacement (cu.in.) & disp & -0.030 & 0.327 & -0.666 & 0.565 & 0.245 & -0.021 & 0.325 & -0.661 & 0.570 & 0.253\\
Gross horsepower & hp & -0.291 & 0.177 & -0.623 & 0.077 & 0.427 & -0.291 & 0.177 & -0.624 & 0.078 & 0.419 \\
Rear axle ratio & drat &0.117 & 0.141 & -0.161 & -0.570 & 0.398 & 0.117 & 0.142 & -0.162 &0.397 & 0.168 \\
Weight (1000 lbs) & wt & -0.569 & 0.180 & -0.919 & -0.214 & 0.932 & -0.570 & 0.180 & -0.919 & -0.213 & 0.938\\
1/4 mile time & qsec & 0.237 & 0.161 & -0.112 & 0.524 & 0.350 & 0.238 & 0.161 & -0.113 & 0.524 & 0.352\\
Straight engine (indicator) & vs & 0.077 & 0.164 & -0.253 & 0.388 & 0.167 & 0.077 & 0.163 & -0.253 & 0.386 & 0.158 \\
Manual transmission (indicator) & am & 0.218 & 0.152 & -0.086 & 0.513 & 0.309 & 0.215 & 0.151 & -0.087 & 0.509 & 0.301\\ 
Number of forward gears & gear & 0.105 & 0.172 & -0.219 & 0.454 & 0.166 & 0.103 & 0.175 & -0.225 & 0.464 & 0.167\\
Number of carburetors & carb & -0.197 & 0.165 & -0.534 & 0.114 & 0.244 & -0.194 & 0.166 & -0.535 & 0.119 & 0.246\\
\end{tabular}
\end{scriptsize}
\end{table}
\addtolength{\tabcolsep}{+0.3em}

Finally, we have included a summary of the five models with the highest posterior probabilities in Table~\ref{tab:res-mtcars-models}. It can be seen how a few variables produce the models with the highest posterior probabilities.

\begin{table}
\caption{Models with the highest posterior probabilities.}
\centering
\begin{tabular}{l|l|c|c}
\multicolumn{2}{c}{}& \multicolumn{2}{|c}{Post. prob.}\\
 \hline
Model &  Variables included & MCMC & Marg. lik. \\
\hline
1000100000 & 1 + wt + cyl & 0.084 & 0.088 \\
0010100000 & 1 + wt + hp & 0.053 & 0.053 \\
1010100000 & 1 + wt + cyl + hp & 0.042 & 0.039 \\
0000110100 & 1 + wt + qsec + am & 0.035 & 0.036 \\
0000110000 & 1 + wt + qsec & 0.034 & 0.035 \\
\end{tabular}
\label{tab:res-mtcars-models}
\end{table}

\subsection{Changepoint model}
\label{subsec:changepoint}

The analysis of time series often requires flexible models that take different structures in different time intervals. For example, change-point models \citep{Carlinetal:1992} can be used to propose models with a structure that changes at particular points in time (i.e., the changepoints). Hence, models can be defined to have a particular structure between any two consecutive change points.

The problem of determining the actual changepoints can be regarded as a complex inferential problem and it can be framed within the problem of model selection.
In particular, models will be defined by the actual changepoints, i.e., the time points at which the model changes from one structure to another. In this regard, the posterior probabilities of being a changepoint could be computed by means of Bayesian inference. It is worth noting that, conditional on the changepoints, models can be written as separate models defined at different non-overlapping time intervals. This means that they can be fit independently of each other.

In order to illustrate model fitting of changepoint models with the approach proposed in this paper, the Great Britain coal mine disasters data will be used. This dataset was first analyzed in \cite{Maguireetal:1952} and then updated and corrected in \cite{Jarrett:1979}. A Bayesian analysis of the dataset has been conducted in \cite{Carlinetal:1992}. 
The dataset records the number of yearly coal-mining disasters (i.e., explosions in which 10 or more miners were killed) between 1851 and 1962.

\begin{figure}
\includegraphics{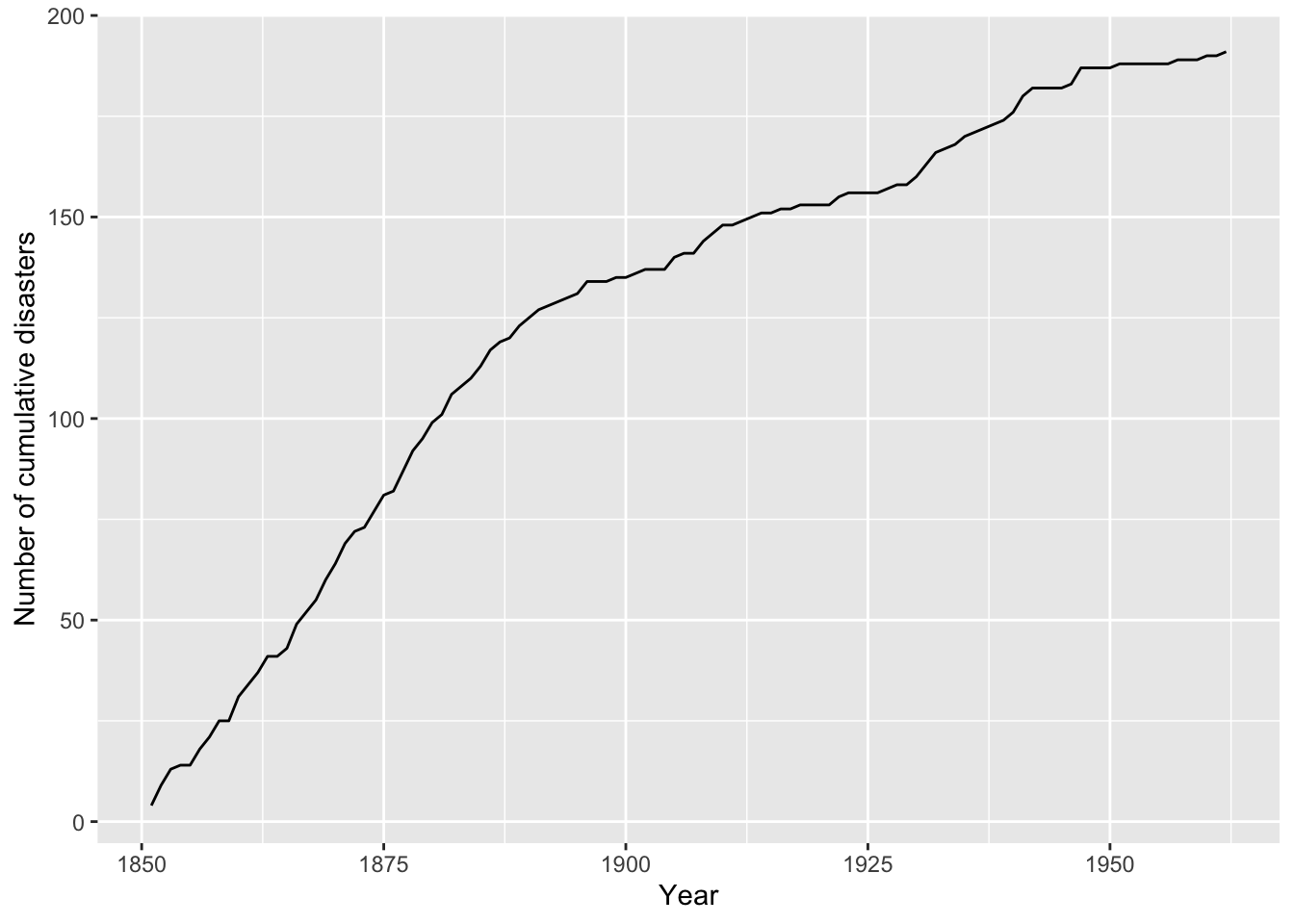}
\caption{Cumulative number of coal-mining accidents in Great Britain between 1851 and 1962.}
\label{fig:coal}
\end{figure}

Figure~\ref{fig:coal} shows the number of cumulative number of accidents, which shows a change in trend by year 1890 and, possibly, another change by year 1945. Hence, it seems plausible to model the yearly number of disasters as a changepoint model with an unknown changepoint.

In particular, the number of coal-mining accidents, $O_t$, in year $t$ can be modeled as as Poisson distribution. Change-point is assumed to be $t_1$, with $1851 \leq t_1 \leq 1962$.
Model is defined as follows:

$$
\begin{array}{cccc}
O_t & \sim & \mathrm{Po}(\mu_1) & t=1851, \ldots, t_1\nonumber\\
\log(\mu_1) & = & \alpha_1 \nonumber\\
O_t & \sim & \mathrm{Po}(\mu_2) & t=t_1+1,\ldots, 1962\nonumber\\
\log(\mu_2) & = & \alpha_2 \nonumber\\
\label{eq:cp-model}
\end{array}
$$
\noindent
Note that when $t_1=1962$ is a situation of no changepoint. Hence, the second component of the model will be missing.

In order to complete the model, period-specific intercepts $\alpha_1$ and $\alpha_2$ are assigned a Normal with zero mean and precision 0.001. Similarly, change point $t_1$, which takes values in $\Omega_t = \{1851, \ldots, 1962\}$ is also assigned a uniform prior so that, a priori, all times in $\Omega_t$ have the same probabilities of being a changepoint.

Note that conditional on $t_1$, the model can be easily split into two separate models for times in $[1851,t_1]$ and $[t_1 +1,1962]$, respectively. These models are straightforward to fit with INLA by considering two different intercepts. However, inference about the changepoint $t_1$ requires a different approach.

We have implemented a Metropolis-Hastings approach as described in Section~\ref{sec:MCMC}. In particular, the proposal distribution for the changepoint takes the current value of 
the changepoint, $t_1^*$, and proposes a new value using a discrete uniform distribution in the set $\{t_1^*-k, \ldots, t_1^*-k+1\}$ with $k$ a non-negative integer. We have taken $k=10$. Note that this set may be truncated when the value of  $t_1^*$ is close to either 1851 or 1962 and the probabilities associated to the proposal distribution must be modified accordingly when computing the acceptance probability.

\begin{figure}
\includegraphics{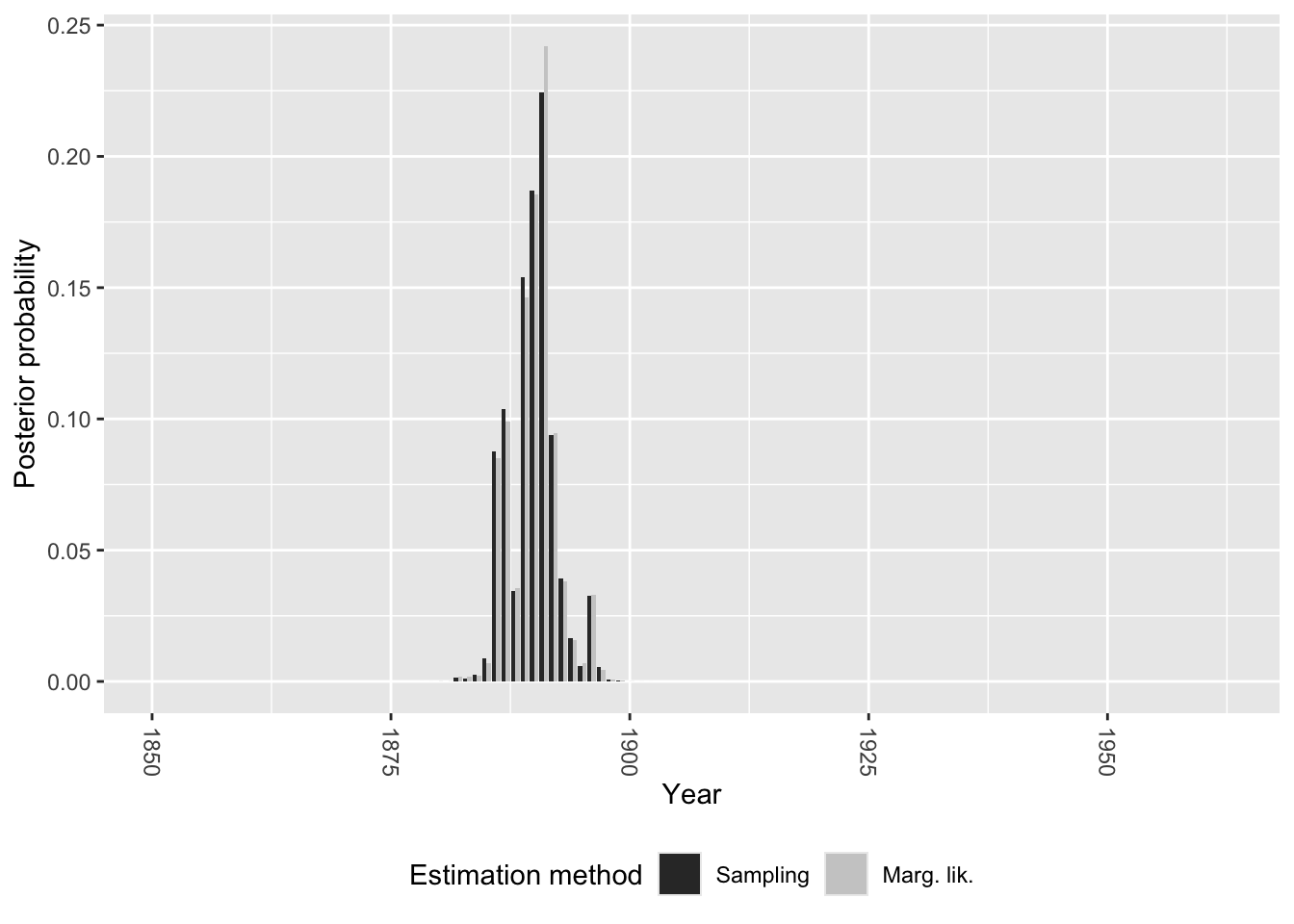}
\caption{Posterior probabilities of the change-point model.}
\label{fig:res-cpmodel}
\end{figure}

We have run the simulations with a starting value of the changepoint of 1906 (the midpoint in the time series). The number of total simulations have been 100000, of which 1000 have been used as a burn-in and for the remaining samples we have kept one in 10 to achieve a total number of 9900 samples. The posterior probabilities of $t_1$ have been computed from the samples as well fitting all models for $t_1=1851,\ldots,1962$ and computing the respective posterior probabilities. Figure~\ref{fig:res-cpmodel} illustrates the posterior probabilities computed by the different methods. As can be seen, all methods provide very close estimates. The year with the highest posterior probability is 1891, with a posterior probability of 0.233 (using MCMC output) and 0.234 (using the marginal likelihoods directly).

Similarly, Table~\ref{tab:res-cpmodel} provides a summary of the model parameters obtained from their respective marginal distributions. Note that the marginal of the different models parameters have been obtained by computing the Bayesian model average of the different methods fitted to the data using the posterior probability of each model as weights. In all cases, estimates found are very similar to those reported in \cite{Carlinetal:1992}.

\begin{table}
\caption{Summary of models parameters estimated with Bayesian model averaging.}
\label{tab:res-cpmodel}
\centering
\begin{tabular}{l|cccc|cccc|}
& \multicolumn{4}{c}{MCMC} & \multicolumn{4}{|c|}{Marg. lik.} \\
\cline{2-9}
Variable  & Mean & St. dev. & \multicolumn{2}{c|}{95\% C.I.} &
Mean & St. dev. & \multicolumn{2}{c|}{95\% C.I.} \\ 
\hline
$\mu_1$ & 3.120 & 0.290 & 2.589  & 3.730 & 3.120 & 0.290 & 2.589  & 3.729\\
\hline
$\mu_2$ & 0.923 & 0.117 & 0.713 & 1.171 & 0.923 & 0.117 & 0.713 & 1.170 \\
\end{tabular}
\end{table}%

%

\subsection{Exposure to pollution sources}
\label{subsec:example3}

\cite{PalmiPeralesetal:2021} study the spatial distribution of cases of three different types of cancer in the Alcal\'a de Henares (Madrid, Spain) to assess an increased risk around different pollution sources. They assess different models to account for spatial variation, socio-economic factors and the distance to the different pollution sources using log-Gaussian Cox Processes (LGCP).

In this particular context, model selection is important to assess association of an increase of cancer cases in the vicinity of several pollution sources. Up to 13 polluting industries have been identified in Alcalá de Henares.

The model proposed is an inhomogeneous Poisson process (IPP) for the cases with intensity $\lambda_0$ and also an IPP for the cases with intensity $\lambda_1(x)$.

$$
\log(\lambda_0) = \alpha_0 + S(x)
$$

$$
\log(\lambda_1) = \alpha_1 + S(x) + \sum_{i=1}^K \beta_i x^{(i)}(x) + \sum_{f\in\mathcal{F}} e_f(x)
$$
\noindent
Here, $\alpha_0$ and $\alpha_1$ are two intercepts (to account for the different number of events), $\beta_i$ is the coefficient of covariate $x^{(i)}(x)$, $S(x)$ is a shared spatial process that will be modeled using a Gausian process with Matérn covariance function \citep[see][for details]{PalmiPeralesetal:2021}  and $e_f(x)$ is a smooth function that measures exposure to pollution source $f$.

Covariates included in the model are unemployment rate (between 20 and 59 years), a score for social class, a measure of education level (between 30 and 39 years) and the percentage of children aged between 0 and 3 that are in the school \citep[see][for more details]{PalmiPeralesetal:2021}. They are included in all models to account for socio-economic risk factors. However, model selection will be on the smooth functions associated to exposure to the pollution sources. In this way, we can assess whether there is an increase of cases around any of the pollution sources 

Regarding the estimation of the marginal likelihood with INLA for LGCP models, \cite{Doversetal:2023} provide a comparison of different methods and they indicate that INLA can provide accurate estimates. Hence, we will rely on the estimates produced by INLA when performing the model selection procedure.

Although there are 13 different pollution sources, only models up to 5 sources are considered. This reduces the set of all possible models from 8192 to 2380, which is still large. This is done because it is nearly impossible to measure the effect of more than 5 pollution sources and, most importantly, given the spatial nature of the effects added, having too many pollution sources in the model will likely cause estimation problems due to identifiability issues.

The method described in Section~\ref{subsec:GS} has been employed to explore the model space. The algorithm has been run for 1000 burn-in interations followed by 1 million iterations, of which one in 10 has been kept, leading to a total of 100000 iterations. Both priors, $\pi_1(\gamma)$ and $\pi_2(\gamma)$, have been considered in order to assess their impact on model selection. 

First of all, the percentage of models explored has been 89.20\% (2123 out of 2380) with $\pi_1(\gamma)$ and 80.67\% (1920 out of 2380) with $\pi_2(\gamma)$. This ensures that the model space is thoroughly explored for a large number of iterations. Note that only
547 and 295 models have been actually selected (i.e., have a count larger than zero) during the sampling process, respectively. Furthermore, when assessing how many models are relevant, 76 and 51 models have a posterior probability greater than 1 / 2380, respectively. We have taken 1 / 2380 as a reasonable baseline posterior probability for the scenario that all models are equally likely. Finally, the minimum number of models to reach a 0.95 posterior probability is 47 and 23, respectively.

\begin{figure}
\centering
\includegraphics[width=10cm]{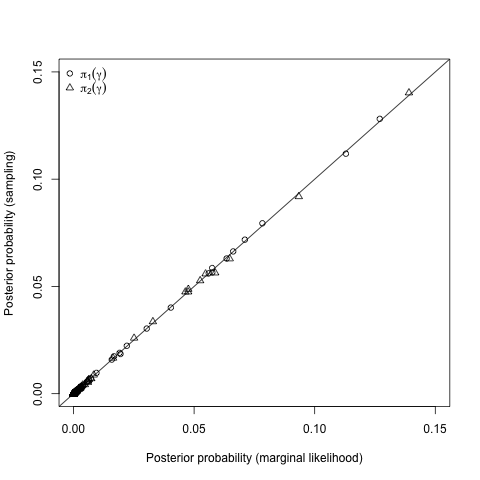}
\caption{Posterior probabilities of the example on exposure to pollution sources. Results show the posterior probabilities obtained with sampling and the marginal likelihoods for the two different priors considered for the models.}
\label{fig:spatial-pprob}
\end{figure}

It is likely that with fewer iterations similar posterior probabilities will be obtained for the most likely models. Furthermore, when the prior includes a higher penalty (i.e., lower prior probability) on the number of components, the percentage of models explored is smaller due to their smaller posterior probabilities. Figure~\ref{fig:spatial-pprob} shows the estimates of the posterior probabilities computed from sampling and the marginal probabilities, and agreement is high between the two methods for both priors.

Table~\ref{tab:GS} shows the posterior probabilities of the 10 most likely models and the sources included, which are the same for both priors (but in a slightly different order). Similarly, Table~\ref{tab:GS2} shows the inclusion probabilities of the different pollution sources. It is worth mentioning that while source 5 is included alone in the most likely model, source 10 is the one with the highest inclusion probability and this happens for both priors.

\begin{table}
\caption{Models with the highest posterior probabilities in the example on exposure to pollution sources.}
\centering
\begin{tabular}{l|c|c|c|c|c}
\multicolumn{2}{c}{}& \multicolumn{2}{|c}{Post. prob. ($\pi_1(\gamma)$)} &
  \multicolumn{2}{|c}{Post. prob. ($\pi_2(\gamma)$)}\\
 \hline
Model &  Sources included &  Marg. lik. & MCMC &  Marg. lik. & MCMC \\
\hline
0000100000000 & 5 & 0.13  & 0.13 &  0.22 & 0.22 \\
0000000001010 & 10, 12 & 0.11 & 0.11 & 0.09 & 0.09\\
0001000001000 & 4, 10 & 0.08 & 0.08 & 0.06 & 0.06\\
0000000001001 & 10, 13 & 0.07 & 0.07 & 0.06 & 0.06\\
0010000001000 & 3, 10 & 0.07 & 0.07 & 0.05 & 0.06\\
0100000001000 & 2, 10 & 0.06 & 0.06 & 0.05 & 0.05\\
1000000001000 & 1, 10 & 0.06 & 0.06 & 0.05 & 0.05\\
0000000101000 & 8, 10 & 0.06 & 0.06 & 0.05 & 0.05\\
0000001001000 & 7, 10 & 0.06 & 0.06 & 0.05 & 0.05\\
0000000000000 & -- & 0.04 & 0.04 & 0.14 & 0.14\\
\end{tabular}
\label{tab:GS}
\end{table}

\begin{table}
\caption{Posterior probabilities of inclusion of the different pollution sources.}
\centering
\begin{tabular}{c|c|c|c|c}
 & \multicolumn{2}{|c|}{Inclusion prob. ($\pi_1(\gamma)$)} &
  \multicolumn{2}{c}{Inclusion prob. ($\pi_2(\gamma)$)}\\
 \hline
Source &    Marg. lik. & MCMC &  Marg. lik. & MCMC \\
\hline
1 & 0.07 & 0.07 & 0.06 & 0.06\\
2 & 0.12 & 0.12 & 0.08 & 0.08\\
3 & 0.10 & 0.10 & 0.09 & 0.10\\
4 &  0.14 & 0.15  & 0.10  & 0.09\\
5 &  0.25 & 0.25  & 0.26 & 0.26\\
6 &  0.05 & 0.05 & 0.04 & 0.04\\
7 &  0.11 & 0.11  & 0.07 & 0.07\\
8 &  0.06 & 0.07 & 0.05 & 0.05\\
9 &  0.03 &  0.03 & 0.01 & 0.01\\
10&  0.73 & 0.73 & 0.54 & 0.54\\
11&  0.01 & 0.01 & 0.01 & 0.01\\
12&  0.14 & 0.14  & 0.11 & 0.11\\
13&  0.12 & 0.12 & 0.08 & 0.08\\
\end{tabular}
\label{tab:GS2}
\end{table}

\section{Discussion and final remarks}
\label{sec:discussion}

This paper illustrates a novel computational approach for model and variable selection that relies on the integrated nested Laplace approximation for fast computation. The approach is general enough as to being able to tackle a large family of models with minor changes in the implementation.

The results and examples presented here confirm that the proposed is a valid approach for model selection that can be employed in a number of contexts. In particular, given the flexibility of INLA to fit mixed-effects models, this approach can be exploited for model and variable selection for spatio-temporal models, survival and joint models and many others.

As illustrated in the examples provided, priors can be used to define how models are penalized by their complexity and other types of priors not mentioned here could be considered. These priors will also define how the model space is explored.

Similarly, other approaches to conduct the exploration of the model space could be proposed. For example, since INLA can fit models very fast, models with a single variable could be fitted even when the number of predictors is very large. Then, the marginal likelihood of each model could be used to control the probability of proposing each variable to enter the model, based on the idea that predictors that do not perform well alone will not do so when other variables are included. This can also be used as a previous step to filter the variables that are considered in the model selection so that variables that are not likely to perform well are not considered. Similarly, before starting the sampling process, some models could be pre-fit in parallel so that these are kept to speed up sampling.

\section{Data Availability Statement}

The \texttt{mtcars} dataset available in the R software for statistical computing. The Great Britain coal mine disasters dataset is publicly available from different sources (including the original paper). The dataset used in the example on exposure to pollution sources is not publicly available due to confidentiality constraints. 

\bibliography{rjinla}
\end{document}